\documentclass[a4paper,11pt]{article}
\usepackage{pos,wrapfig,enumitem,xspace}

\newcommand{\eV}{\mathrm{EeV}}
\newcommand{\EeV}{\mathrm{EeV}}
\newcommand{\Mpc}{\mathrm{Mpc}}
\newcommand{\km}{\mathrm{km}}
\newcommand{\yr}{\mathrm{yr}}
\newcommand{\sr}{\mathrm{sr}}
\providecommand{\degree}{^\circ}

\newcommand{\dO}{\operatorname{d\Omega}}
\newcommand{\skyint}[1]{\int_{4\pi} #1 \dO}

\usepackage{lineno}

\title{UHECR arrival directions in the latest data from the original Auger and TA surface detectors and nearby galaxies}
\ShortTitle{UHECR arrival directions and nearby galaxies}
\author*[a]{Armando di Matteo}
\author{Luis Anchordoqui}\author{Teresa Bister}\author{Jonathan Biteau}\author{Lorenzo~Caccianiga}\author{Rogério de~Almeida}\author{Olivier Deligny}\author{Ugo Giaccari}\author{Diego~Harari}\author{Jihyun Kim}\author{Mikhail Kuznetsov}\author{Ioana Mariș}\author{Grigory Rubtsov}\author{Peter~Tinyakov}\author{Sergey Troitsky}\author{Federico Urban}
\affiliation[a]{Istituto Nazionale di Fisica Nucleare (INFN), sezione di Torino,
Via Pietro Giuria 1, 10125 Torino, Italy}
\affiliation[b]{Observatorio Pierre Auger, Av.\ San Mart{\'\i}n Norte 304, 5613 Malarg\"ue, Argentina
}
\affiliation[c]{Telescope Array Project, 201 James Fletcher Bldg, 115 S.\ 1400 East, Salt Lake City, UT 84112-0830, USA
}
\forColl{Pierre Auger$^b$}\forColl{ and the Telescope Array$^c$}
\emailAdd{spokespersons@auger.org}
\emailAdd{ta-icrc@cosmic.utah.edu}

\abstract{The distribution of ultra-high-energy cosmic-ray arrival directions appears to be nearly isotropic except for a dipole moment of order $6 \times (E/10~\EeV)$~per~cent. Nonetheless, at the highest energies, as the number of possible candidate sources within the propagation horizon and the magnetic deflections both shrink, smaller-scale anisotropies might be expected to emerge.  On the other hand, the flux suppression reduces the statistics available for searching for such anisotropies.  In this work, we consider two different lists of candidate sources: a sample of nearby starburst galaxies and the 2MRS catalog tracing stellar mass within~$250~\Mpc$.\\We combine surface-detector data collected at the Pierre Auger Observatory until 2020 and the Telescope Array until 2019, and use them to test models in which UHECRs comprise an isotropic background and a foreground originating from the candidate sources and randomly deflected by magnetic fields. The free parameters of these models are the energy threshold, the signal fraction, and the search angular scale. 
We find a correlation between the arrival directions of $11.8\%_{-3.1\%}^{+5.0\%}$~of cosmic rays detected with $E \ge 38~\EeV$ by~Auger or with~$E \gtrsim 49~\EeV$ by~TA and the position of nearby starburst galaxies on a ${15.5\degree}_{-3.2\degree}^{+5.3\degree}$~angular scale, with a $4.2\sigma$~post-trial significance, as well as a weaker correlation with the overall galaxy distribution.}

\FullConference{37$^{\rm{th}}$ International Cosmic Ray Conference (ICRC 2021)\\
		July 12th -- 23rd, 2021\\
		Online -- Berlin, Germany}

\begin{document}
\maketitle

\section{Introduction}
Ultra-high-energy cosmic rays (UHECRs) are particles from outer space with energies greater than $1~\EeV = 10^{18}~\eV \approx 0.16~\mathrm{J}$.
They are electrically charged (protons and other atomic nuclei), so they are deflected by intergalactic and Galactic magnetic fields (typically by a few tens of degrees), meaning that, unlike with photons and other neutral messengers, the position of their sources cannot be directly reconstructed from their arrival directions.

Nowadays, arrays of particle detectors such as the Pierre Auger Observatory (Auger) \cite{bib:Auger} and the Telescope Array (TA) \cite{bib:TA} cover areas of hundreds of square kilometers and detect thousands of events every year; nevertheless, over 60 years after the discovery of UHECRs, their origin remains un\-known.  
Still, certain possibilities can be ruled out.  
The lack of anisotropies aligned with the Galactic plane excludes a sizable contribution of protons from within our Galaxy \cite{bib:TA-nogal,bib:Auger-nogal}, and mass estimates exclude a composition dominated by heavier nuclei \cite{bib:Auger-mass,bib:TA-mass}, hence most such particles must originate from outside our Galaxy.
The lack of neutral particles such as neutrinos and gamma rays at these energies \cite{bib:Auger-nu,bib:TA-nu,bib:Auger-gamma,bib:TA-gamma} excludes ``top-down'' mechanisms, e.g.\ the decay of super-heavy dark matter particles or topological defects, as a dominant origin (except possibly at~$E \gtrsim 100~\EeV$).
Therefore, UHECRs are widely believed to be ordinary matter accelerated to extreme energies by extragalactic astrophysical phenomena.  Various possibilities that have been hypothesized \cite{bib:review2,bib:review} include active galactic nuclei (AGNs), starburst galaxies (SBGs), gamma-ray bursts (GRBs) and tidal disruption events (TDEs).

A possible avenue to search for imprints of the distribution of UHECR sources in spite of magnetic deflections is to harness the huge statistics gathered by last-generation detector arrays to search for large-scale (dipolar and quadrupolar) anisotropies, which are the ones the least affected by a given amount of magnetic deflections.
Another way is to focus on the highest-energy part of the UHECR spectrum, where magnetic deflections are expected to be smaller and the number of potential sources decreases, at the cost of the reduced statistics.
A large-scale anisotropy has been reported in Auger data \cite{bib:AugerScience2017} whose statistical significance has now reached~$6.6\sigma$ \cite{bib:AugerLSA2021}, but the lack of full-sky coverage impedes its interpretation in terms of dipole and quadrupole moments unless higher-order multipoles are assumed to vanish.
Conversely, no medium- or small-scale anisotropy has been conclusively established so far, but a few indications have been reported (see the introduction of Ref.~\cite{bib:MSA2019} for a review).
In order to follow up on these indications using full-sky data, a working group has been established with members from both the Auger and TA collaborations. Our most recent results of searches for large-scale anisotropies are presented in Ref.~\cite{bib:LSA2021}, and those of searches for medium-scale anisotropies at the highest energies are presented here.

\section{The datasets}
In this work, we use the same data as in Ref.~\cite{bib:LSA2021}, namely those detected by the Pierre Auger Observatory from~2004 Jan~01 to~2020 Dec~31 and those detected by the Telescope Array from~2008 May~11 to~2019 May~10, but restricted to the highest-energy bin (above~$32~\EeV$ for Auger, above~$40.8~\EeV$ for TA), and using looser selection criteria for Auger events \cite{bib:AugerMSA2021} resulting in $7\%$~more events.  The dataset comprises $2\,625$ Auger events and 315 TA events.

The geometrical exposure is $95\,700~\km^2~\yr~\sr$ for Auger vertical events (zenith angles~$\theta < 60\degree$) and $26\,300~\km^2~\yr~\sr$ for Auger inclined events ($60\degree \le \theta < 80\degree$).  
Taking into account the energy resolution effects, the effective exposure is $96\,600~\km^2~\yr~\sr$ for Auger vertical events, $26\,600~\km^2~\yr~\sr$ for Auger inclined events, and $13\,700~\km^2~\yr~\sr$ for TA events.  
This represents a $33\%$~increase from the last Auger--TA joint searches for medium-scale anisotropies \cite{bib:MSA2019}.  
\begin{figure}
    \centering
    \includegraphics[page=4]{Figures/expos_plot.pdf}
    \hfil
    \includegraphics[]{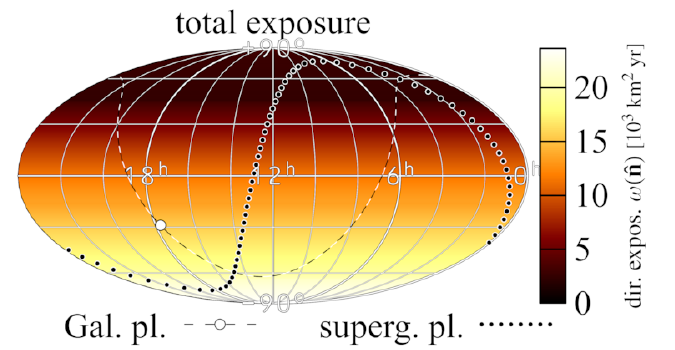}
    \caption{The directional exposure of the datasets we used. The yellow area in the left panel is the fiducial declination band used for the cross-calibration of energies \cite{bib:LSA2021}.}
    \label{fig:exposure}
\end{figure}
The declination dependence of the directional exposure is computed in the approximation of $100\%$~detector efficiency \cite{bib:exposure} and shown in~\autoref{fig:exposure}.

Following Ref.~\cite{bib:LSA2021}, we apply the conversion \begin{equation}
    E_\text{TA} \mapsto E_\text{Auger} = 8.57 \left(E_\text{TA}/{10~\EeV}\right)^{0.937}~\EeV
\end{equation} to TA event energies in order to correct them for the mismatch in the energy scales of the two experiments, which has been estimated by comparing their data in a common declination band in the intersection of their fields of view.
\begin{figure}
    \centering
    \includegraphics{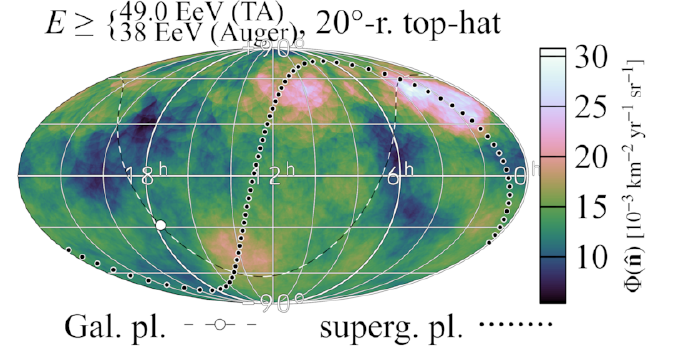}
    \hfil
    \includegraphics{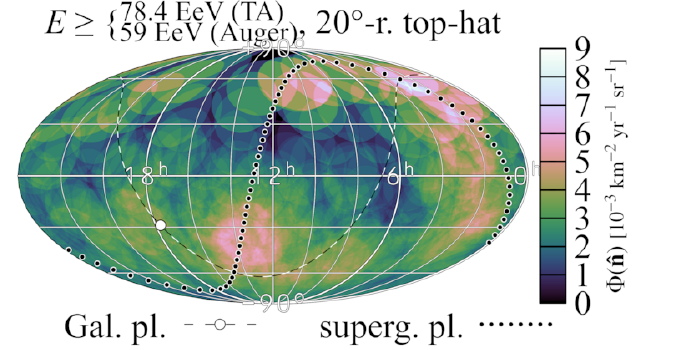}
    \caption{The flux distribution from our dataset above two selected energy thresholds, in equatorial coordinates}
    \label{fig:tophat_maps}
\end{figure}
The distribution of arrival directions of the events above two selected energy thresholds, averaged over $20\degree$-radius top-hat windows, is shown in \autoref{fig:tophat_maps}.

\section{The analysis}

\newcommand{\n}{\mathbf{\hat n}}
\newcommand{\TS}{\mathrm{TS}}
In this work, we present the result of a likelihood ratio test between 
flux models including a contribution from nearby galaxies and
and the isotropic null hypothesis, similar to Refs.~\cite{bib:AugerSBG,bib:TASBG,bib:AugerMSA2021}.

We define the test statistic \begin{align}
    \TS(\psi,f,E_{\min}) &= 2 \ln \frac {L(\psi,f,E_{\min})}{L(\psi,0,E_{\min})},
    & L(\psi,f,E_{\min}) &= \prod_{E_i \ge E_{\min}} \frac{\Phi(\n_i;\psi,f)\omega(\n_i)}{\skyint{\Phi(\n;\psi,f)\omega(\n)}},
\end{align} where $\omega(\n)$~is the combined directional exposure of the dataset, and the flux model is \begin{equation}
    \Phi(\n; \psi, f) = f \Phi_\text{signal}(\n; \psi) + (1-f) \Phi_\text{background},
\end{equation} where the contribution of each source is a von Mises--Fisher distribution: \begin{align}
    \Phi_\text{signal}(\n; \psi) &= 
    \frac{1}{\sum_j w_s}
    \sum_j w_s \frac{\psi^{-2}}{4\pi\sinh\psi^{-2}} \exp\left(\psi^{-2}\n_s \cdot \n\right); & \Phi_\text{background} &= \frac{1}{4\pi},
\end{align}
where
$E_i$~and $\n_i$~are the energy and arrival direction of the $i$-th event; $w_s$~and $\n_s$~are the weight and position of the $s$-th source candidate as defined in \autoref{sec:catalogs};
and $\psi$~is the root-mean-square deflection per transverse dimension (i.e.\ the total r.m.s.\ deflection is~$\sqrt{2}\times\psi$).\footnote{The equivalent top-hat radius is $\Psi = 1.59\psi$ \cite{bib:AugerMSA2019}.}  The von Mises--Fisher distribution is the analog of a Gaussian on a 2-sphere, centered on the position of each source.  In reality, magnetic deflections include both regular and turbulent parts, but the directions of the former are not sufficiently well known to be used in a log-likelihood ratio test.  In a future work, we plan to investigate using simulations how much realistic amounts of regular magnetic deflections can affect the result of an analysis which does not explicitly model them.

Since the null hypothesis (isotropy) is a special case of the model (obtained for~$f=0$) and for a fixed~$E_{\min}$ the $\TS$~is a smooth function of~$\psi$ and~$f$, according to Wilks' theorem \cite{bib:Wilks} $\max_{\psi,f}\TS$ is $\chi^2$-distributed with two degrees of freedom.

The analysis is repeated using energy thresholds of~$32~\EeV, 33~\EeV, \ldots, 80~\EeV$ on the Auger scale, corresponding to~$40.8~\EeV, 42.2~\EeV, \ldots, 108.6~\EeV$ on the TA scale.

\subsection{The galaxy catalogs}
\label{sec:catalogs}
In this work, we use two different lists of candidate sources.
The first is a list of~$44\,113$ galaxies of all types at distances~$1~\Mpc \le D < 250~\Mpc$, based on the 2MASS catalog with distances from HyperLEDA, with weights assumed proportional to the near-infrared flux in the $K$-band ($2.2~\mathrm{\mu m}$).
The second is a list of 44 starburst galaxies at distances~$1~\Mpc \le D < 130~\Mpc$, taken from Ref.~\cite{bib:Lunardini2019} except that we removed the SMC and LMC (which are dwarf irregular galaxies, not starburst galaxies, as evidenced by their infrared-to-radio flux ratio much lower than all other objects of the list), and added the Circinus galaxy with data from the Parkes telescope ($\alpha=213.29\degree$, $\delta=-65.34\degree$, $D=4
.21$, $S_{1.4~\mathrm{GHz}}=1.50~\mathrm{Jy}$); these galaxies were assigned weights proportional to their radio flux at~$1.4~\mathrm{GHz}$.  More details about these selections are found in Ref.~\cite{bib:AugerMSA2021}.

In this work, we neglect the energy losses undergone by cosmic rays, hence the distant objects are assigned a larger weight than if energy losses were taken into account.  Given the distance distributions of the objects we are considering, the effect of energy losses on the results can be presumed to be relatively small in the case of galaxies of all types and  negligible in the case of starburst galaxies, though the precise rates depend on the mass composition of UHECRs.
Also, like in previous studies \cite{bib:AugerSBG,bib:TASBG,bib:AugerMSA2021}, we do not attempt to correct for the fact that the catalogs are limited in flux rather than in intrinsic luminosity, meaning that distant objects can be excluded even if otherwise-identical objects would be included if closer to us.  An estimate of the size of the effects of this limitation on the results is left for future works.

\section{Results}

Using the list of starburst galaxies, we find a maximum test statistic of~$\TS=27.2$ with an energy threshold of~$E_{\min}=38~\EeV$ on the Auger scale ($49~\EeV$ on the TA scale), an angular scale~$\psi={15.5\degree}_{-3.2\degree}^{+5.3\degree}$,\footnote{Equivalent top-hat radius: $\Psi = {24.6\degree}_{-5.1\degree}^{+8.4\degree}$.} and a signal fraction~$f=11.8\%_{-3.1\%}^{+5.0\%}$.  Using the list of all types of galaxies galaxies, we find a maximum~$\TS=16.2$ with~$E_{\min}=41~\EeV$ on the Auger scale ($53~\EeV$ on the TA scale), $\psi={24\degree}_{-8\degree}^{+13\degree}$,\footnote{Equivalent top-hat radius: $\Psi = {38\degree}_{-13\degree}^{+21\degree}$.} and~$f=38\%_{-14\%}^{+28\%}$.  
\begin{figure}
    \centering
    \includegraphics{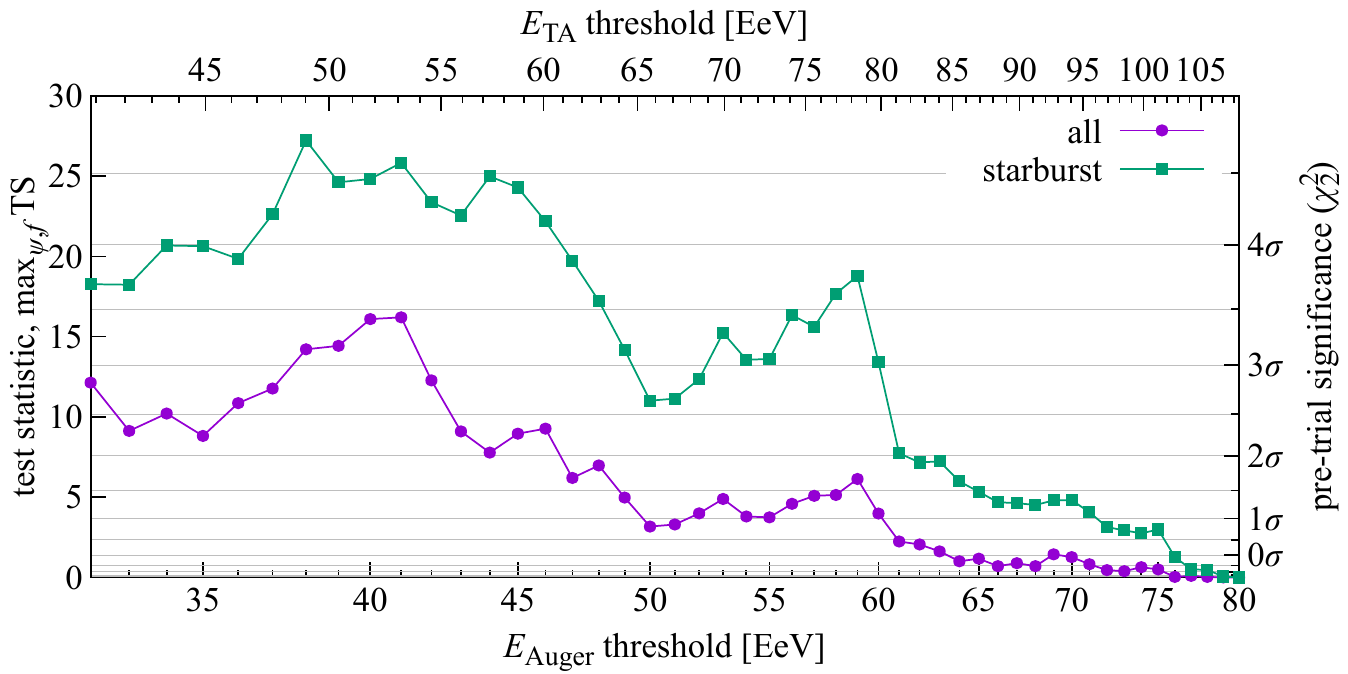}
    \includegraphics{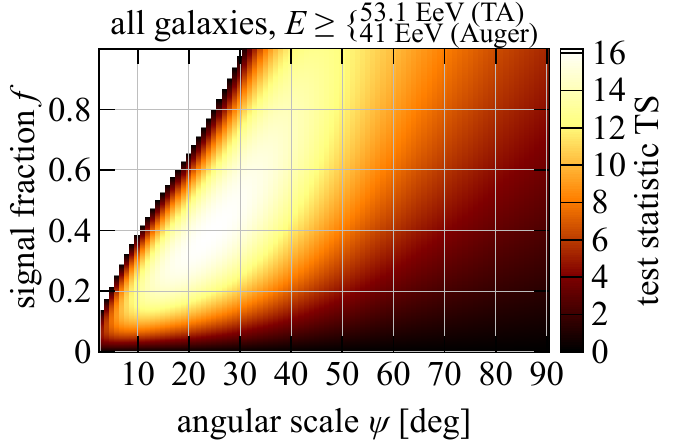}
    \hfil
    \includegraphics{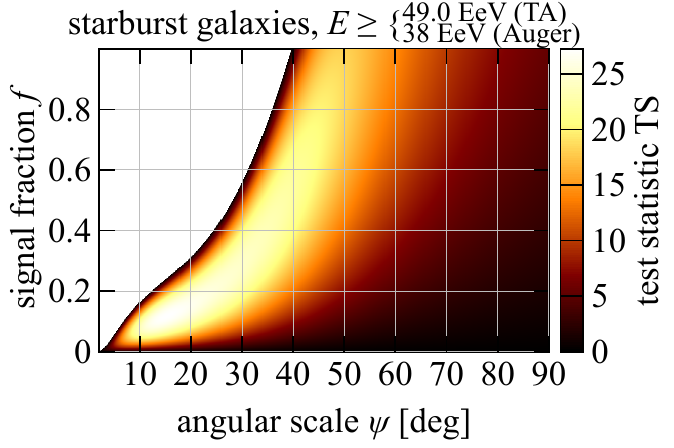}
    \caption{The test statistic~$\TS$ as a function of the energy threshold~$E_{\min}$ (top) and of the angular scale~$\psi$ and signal fraction~$f$ (bottom).  In the top panel, for each~$E_{\min}$ the corresponding best~$\psi$ and~$f$ are used, whereas in each of the bottom panels the same~$E_{\min}$ is used for all~$\psi$ and~$f$.  In the top panel, the significances on the right-hand side take into account the maximization over~$\psi$ and~$f$ but not over~$E_{\min}$. The white areas at small~$\psi$ in the bottom panels correspond to models with~$\TS<0$, i.e.\ fitting the data worse than the isotropic null hypothesis.}
    \label{fig:ts}
\end{figure}
The~$\TS$ as a function of the parameters for the two catalogs is shown in \autoref{fig:ts}.
\begin{figure}
    \centering
    \includegraphics{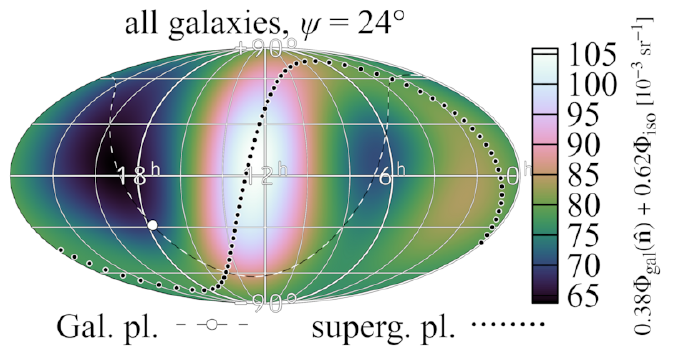}
    \hfil
    \includegraphics{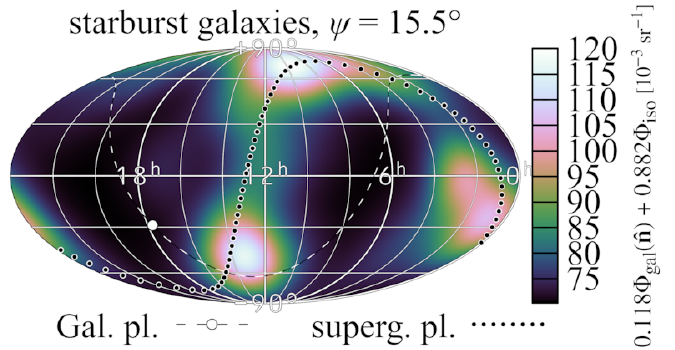}
    \caption{The best-fit flux models for the two catalogs we used}
    \label{fig:model_maps}
\end{figure}
The best-fit flux models are shown in \autoref{fig:model_maps}.
\begin{figure}
    \centering
    \includegraphics[page=2]{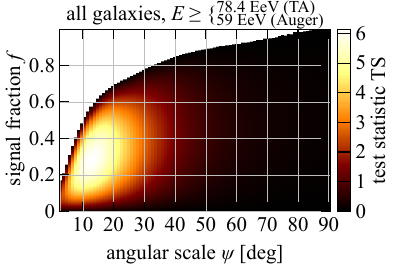}
    \hfil
    \includegraphics[page=2]{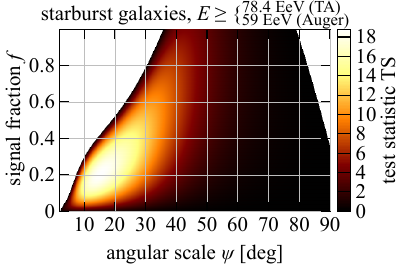}
    \caption{Same as in \autoref{fig:ts}~bottom, but with a higher energy threshold}
    \label{fig:ts_59}
\end{figure}
Using both catalogs, there also is a local maximum at~$E_{\min} = 59~\EeV$ on the Auger scale ($78~\EeV$ on the TA scale); the~$\TS$ as a function of~$\psi$ and~$f$ at this threshold is shown in \autoref{fig:ts_59}.

According to Wilks' theorem \cite{bib:Wilks}, when accounting for the scan over~$\psi$ and~$f$ (but not~$E_{\min}$) these test statistics correspond to local statistical significances of~$4.7\sigma$ and~$3.4\sigma$ respectively.  Wilks' theorem is not applicable to~$E_{\min}$ because the likelihood is not a smooth function of it, so we computed the post-trial significances accounting for all three free parameters using simulations in each of which the number and energies of events are the same as in the real data, but the arrival directions are randomly generated according to the combined directional exposure of the two arrays.  
\begin{figure}
    \centering
    \includegraphics[page=1]{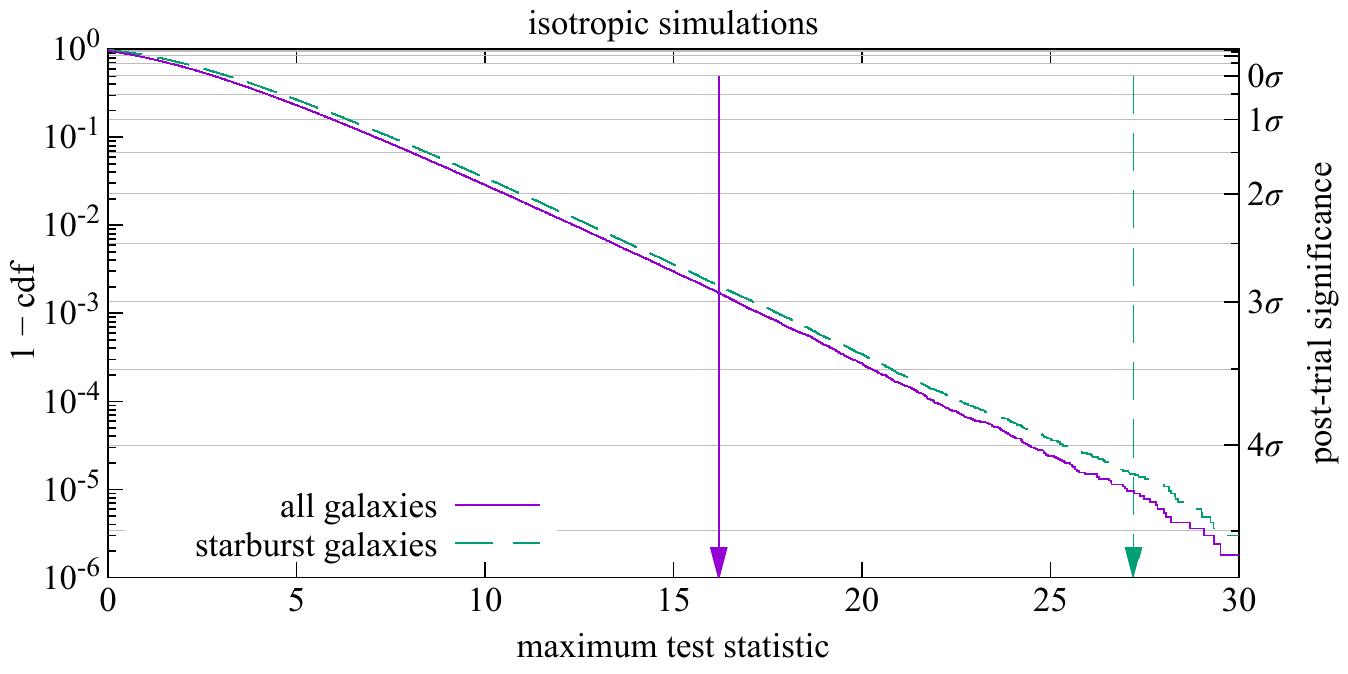}
    \caption{The distribution of $\TS$ among isotropic simulations}
    \label{fig:posttrial}
\end{figure}
The resulting distribution of test statistics is shown in \autoref{fig:posttrial}.  We find that for the starburst galaxy model $\TS=27.2$~corresponds to a $4.2\sigma$~post-trial significance, and for the all-galaxy model $\TS=16.2$~corresponds to a $2.9\sigma$~post-trial significance.

\subsection{Effect of the uncertainty in the energy cross calibration}
As explained in Ref.~\cite{bib:LSA2021}, the statistical uncertainty in the cross calibration of energies can be treated as a $\pm6.4\%$ uncertainty on the ratio between ``effective'' exposures, but we find that such an uncertainty has a negligible effect on the current study: if we increase the TA exposure by~$\pm6.4\%$, the maximum~$\TS$ changes by~$\mp0.4$ and~$\pm 0.1$, for the starburst galaxy model and the all-galaxy model respectively, with changes in~$\psi$ and~$f$ of less than~$1\degree$ and~$1\%$ respectively.  The reason for this is that neither hemisphere dominates the anisotropic component of either model, so the fit to the data cannot be substantially improved or worsened just by rescaling the flux in one or the other hemisphere.

\section{Conclusion}
Our combined dataset hints at an association between the arrival directions of around~$12\%$ of cosmic rays detected with $E \ge 38~\EeV$ by~Auger or with~$E \gtrsim 49~\EeV$ by~TA and the position of nearby starburst galaxies on an angular scale of around~$16\degree$, with a stronger significance than the Auger-only data \cite{bib:AugerMSA2021} but still short of the discovery level, as well as a weaker association with the overall galaxy distribution.  The astrophysical interpretation of this association is complicated by our incomplete knowledge about intergalactic and Galactic magnetic fields and the UHECR mass composition. Therefore we leave the possible interpretations of these results for future studies.

In the coming years, the upgraded arrays AugerPrime~\cite{bib:AugerPrime} and TA$\times$4~\cite{bib:TAx4} will gather more exposure, allowing us to probe flux models with more statistical sensitivity.  It will be interesting to see if the new data will be able to confirm or dispute this finding.  Furthermore, improved mass estimation from new analysis techniques (such as ones involving machine learning \cite{bib:Auger-ML,bib:TA-ML}) and from the new detectors of Auger~\cite{bib:AugerPrime} will allow us to select high-rigidity event samples, which are expected to undergo smaller magnetic deflections.

\newcommand{\etal}{et~al.}
\newcommand{\journal}[5]{\href{https://doi.org/#5}{\textit{#1}\ \textbf{#2} (#3)\ #4}}
\newcommand{\arXiv}[1]{\href{https://arxiv.org/abs/#1}{\nolinkurl{#1}}}

\clearpage
\section*{The Pierre Auger Collaboration}
\small

\begin{wrapfigure}[8]{l}{0.11\linewidth}
\vspace{-5mm}
\includegraphics[width=0.98\linewidth]{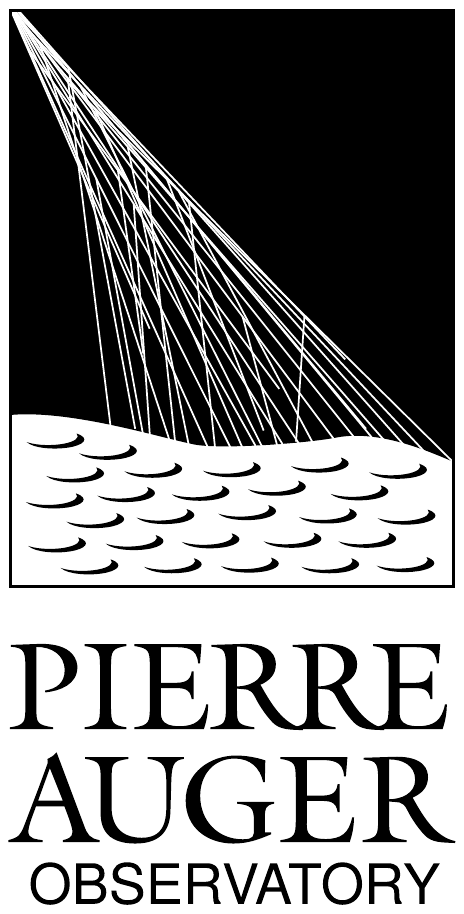}
\end{wrapfigure}
\begin{sloppypar}\noindent
P.~Abreu$^{72}$,
M.~Aglietta$^{54,52}$,
J.M.~Albury$^{13}$,
I.~Allekotte$^{1}$,
A.~Almela$^{8,12}$,
J.~Alvarez-Mu\~niz$^{79}$,
R.~Alves Batista$^{80}$,
G.A.~Anastasi$^{63,52}$,
L.~Anchordoqui$^{87}$,
B.~Andrada$^{8}$
% ...

\end{sloppypar}

\begin{center}
\rule{0.1\columnwidth}{0.5pt}
\raisebox{-0.4ex}{\scriptsize$\bullet$}
\rule{0.1\columnwidth}{0.5pt}
\end{center}

\vspace{-1ex}
\footnotesize

\begin{description}[labelsep=0.2em,align=right,labelwidth=0.7em,labelindent=0em,leftmargin=2em,noitemsep]
\item[$^{1}$] Centro At\'omico Bariloche and Instituto Balseiro (CNEA-UNCuyo-CONICET), San Carlos de Bariloche, Argentina
\item[$^{2}$] Centro de Investigaciones en L\'aseres y Aplicaciones, CITEDEF and CONICET, Villa Martelli, Argentina
\item[$^{3}$] Departamento de F\'\i{}sica and Departamento de Ciencias de la Atm\'osfera y los Oc\'eanos, FCEyN, Universidad de Buenos Aires and CONICET, Buenos Aires, Argentina
\item[$^{4}$] IFLP, Universidad Nacional de La Plata and CONICET, La Plata, Argentina
\item[$^{5}$] Instituto de Astronom\'\i{}a y F\'\i{}sica del Espacio (IAFE, CONICET-UBA), Buenos Aires, Argentina
\item[$^{6}$] Instituto de F\'\i{}sica de Rosario (IFIR) -- CONICET/U.N.R.\ and Facultad de Ciencias Bioqu\'\i{}micas y Farmac\'euticas U.N.R., Rosario, Argentina
\item[$^{7}$] Instituto de Tecnolog\'\i{}as en Detecci\'on y Astropart\'\i{}culas (CNEA, CONICET, UNSAM), and Universidad Tecnol\'ogica Nacional -- Facultad Regional Mendoza (CONICET/CNEA), Mendoza, Argentina
\item[$^{8}$] Instituto de Tecnolog\'\i{}as en Detecci\'on y Astropart\'\i{}culas (CNEA, CONICET, UNSAM), Buenos Aires, Argentina
\item[$^{9}$] International Center of Advanced Studies and Instituto de Ciencias F\'\i{}sicas, ECyT-UNSAM and CONICET, Campus Miguelete -- San Mart\'\i{}n, Buenos Aires, Argentina
\item[$^{10}$] Observatorio Pierre Auger, Malarg\"ue, Argentina
\item[$^{11}$] Observatorio Pierre Auger and Comisi\'on Nacional de Energ\'\i{}a At\'omica, Malarg\"ue, Argentina
\item[$^{12}$] Universidad Tecnol\'ogica Nacional -- Facultad Regional Buenos Aires, Buenos Aires, Argentina
\item[$^{13}$] University of Adelaide, Adelaide, S.A., Australia
\item[$^{14}$] Universit\'e Libre de Bruxelles (ULB), Brussels, Belgium
\item[$^{15}$] Vrije Universiteit Brussels, Brussels, Belgium
\item[$^{16}$] Centro Brasileiro de Pesquisas Fisicas, Rio de Janeiro, RJ, Brazil
\item[$^{17}$] Centro Federal de Educa\c{c}\~ao Tecnol\'ogica Celso Suckow da Fonseca, Nova Friburgo, Brazil
\item[$^{18}$] Instituto Federal de Educa\c{c}\~ao, Ci\^encia e Tecnologia do Rio de Janeiro (IFRJ), Brazil
\item[$^{19}$] Universidade de S\~ao Paulo, Escola de Engenharia de Lorena, Lorena, SP, Brazil
\item[$^{20}$] Universidade de S\~ao Paulo, Instituto de F\'\i{}sica de S\~ao Carlos, S\~ao Carlos, SP, Brazil
\item[$^{21}$] Universidade de S\~ao Paulo, Instituto de F\'\i{}sica, S\~ao Paulo, SP, Brazil
\item[$^{22}$] Universidade Estadual de Campinas, IFGW, Campinas, SP, Brazil
\item[$^{23}$] Universidade Estadual de Feira de Santana, Feira de Santana, Brazil
\item[$^{24}$] Universidade Federal do ABC, Santo Andr\'e, SP, Brazil
\item[$^{25}$] Universidade Federal do Paran\'a, Setor Palotina, Palotina, Brazil
\item[$^{26}$] Universidade Federal do Rio de Janeiro, Instituto de F\'\i{}sica, Rio de Janeiro, RJ, Brazil
\item[$^{27}$] Universidade Federal do Rio de Janeiro (UFRJ), Observat\'orio do Valongo, Rio de Janeiro, RJ, Brazil
\item[$^{28}$] Universidade Federal Fluminense, EEIMVR, Volta Redonda, RJ, Brazil
\item[$^{29}$] Universidad de Medell\'\i{}n, Medell\'\i{}n, Colombia
\item[$^{30}$] Universidad Industrial de Santander, Bucaramanga, Colombia
\item[$^{31}$] Charles University, Faculty of Mathematics and Physics, Institute of Particle and Nuclear Physics, Prague, Czech Republic
\item[$^{32}$] Institute of Physics of the Czech Academy of Sciences, Prague, Czech Republic
\item[$^{33}$] Palacky University, RCPTM, Olomouc, Czech Republic
\item[$^{34}$] CNRS/IN2P3, IJCLab, Universit\'e Paris-Saclay, Orsay, France
\item[$^{35}$] Laboratoire de Physique Nucl\'eaire et de Hautes Energies (LPNHE), Sorbonne Universit\'e, Universit\'e de Paris, CNRS-IN2P3, Paris, France
\item[$^{36}$] Univ.\ Grenoble Alpes, CNRS, Grenoble Institute of Engineering Univ.\ Grenoble Alpes, LPSC-IN2P3, 38000 Grenoble, France
\item[$^{37}$] Universit\'e Paris-Saclay, CNRS/IN2P3, IJCLab, Orsay, France
\item[$^{38}$] Bergische Universit\"at Wuppertal, Department of Physics, Wuppertal, Germany
\item[$^{39}$] Karlsruhe Institute of Technology (KIT), Institute for Experimental Particle Physics, Karlsruhe, Germany
\item[$^{40}$] Karlsruhe Institute of Technology (KIT), Institut f\"ur Prozessdatenverarbeitung und Elektronik, Karlsruhe, Germany
\item[$^{41}$] Karlsruhe Institute of Technology (KIT), Institute for Astroparticle Physics, Karlsruhe, Germany
\item[$^{42}$] RWTH Aachen University, III.\ Physikalisches Institut A, Aachen, Germany
\item[$^{43}$] Universit\"at Hamburg, II.\ Institut f\"ur Theoretische Physik, Hamburg, Germany
\item[$^{44}$] Universit\"at Siegen, Department Physik -- Experimentelle Teilchenphysik, Siegen, Germany
\item[$^{45}$] Gran Sasso Science Institute, L'Aquila, Italy
\item[$^{46}$] INFN Laboratori Nazionali del Gran Sasso, Assergi (L'Aquila), Italy
\item[$^{47}$] INFN, Sezione di Catania, Catania, Italy
\item[$^{48}$] INFN, Sezione di Lecce, Lecce, Italy
\item[$^{49}$] INFN, Sezione di Milano, Milano, Italy
\item[$^{50}$] INFN, Sezione di Napoli, Napoli, Italy
\item[$^{51}$] INFN, Sezione di Roma ``Tor Vergata'', Roma, Italy
\item[$^{52}$] INFN, Sezione di Torino, Torino, Italy
\item[$^{53}$] Istituto di Astrofisica Spaziale e Fisica Cosmica di Palermo (INAF), Palermo, Italy
\item[$^{54}$] Osservatorio Astrofisico di Torino (INAF), Torino, Italy
\item[$^{55}$] Politecnico di Milano, Dipartimento di Scienze e Tecnologie Aerospaziali , Milano, Italy
\item[$^{56}$] Universit\`a del Salento, Dipartimento di Matematica e Fisica ``E.\ De Giorgi'', Lecce, Italy
\item[$^{57}$] Universit\`a dell'Aquila, Dipartimento di Scienze Fisiche e Chimiche, L'Aquila, Italy
\item[$^{58}$] Universit\`a di Catania, Dipartimento di Fisica e Astronomia, Catania, Italy
\item[$^{59}$] Universit\`a di Milano, Dipartimento di Fisica, Milano, Italy
\item[$^{60}$] Universit\`a di Napoli ``Federico II'', Dipartimento di Fisica ``Ettore Pancini'', Napoli, Italy
\item[$^{61}$] Universit\`a di Palermo, Dipartimento di Fisica e Chimica ''E.\ Segr\`e'', Palermo, Italy
\item[$^{62}$] Universit\`a di Roma ``Tor Vergata'', Dipartimento di Fisica, Roma, Italy
\item[$^{63}$] Universit\`a Torino, Dipartimento di Fisica, Torino, Italy
\item[$^{64}$] Benem\'erita Universidad Aut\'onoma de Puebla, Puebla, M\'exico
\item[$^{65}$] Unidad Profesional Interdisciplinaria en Ingenier\'\i{}a y Tecnolog\'\i{}as Avanzadas del Instituto Polit\'ecnico Nacional (UPIITA-IPN), M\'exico, D.F., M\'exico
\item[$^{66}$] Universidad Aut\'onoma de Chiapas, Tuxtla Guti\'errez, Chiapas, M\'exico
\item[$^{67}$] Universidad Michoacana de San Nicol\'as de Hidalgo, Morelia, Michoac\'an, M\'exico
\item[$^{68}$] Universidad Nacional Aut\'onoma de M\'exico, M\'exico, D.F., M\'exico
\item[$^{69}$] Universidad Nacional de San Agustin de Arequipa, Facultad de Ciencias Naturales y Formales, Arequipa, Peru
\item[$^{70}$] Institute of Nuclear Physics PAN, Krakow, Poland
\item[$^{71}$] University of \L{}\'od\'z, Faculty of High-Energy Astrophysics,\L{}\'od\'z, Poland
\item[$^{72}$] Laborat\'orio de Instrumenta\c{c}\~ao e F\'\i{}sica Experimental de Part\'\i{}culas -- LIP and Instituto Superior T\'ecnico -- IST, Universidade de Lisboa -- UL, Lisboa, Portugal
\item[$^{73}$] ``Horia Hulubei'' National Institute for Physics and Nuclear Engineering, Bucharest-Magurele, Romania
\item[$^{74}$] Institute of Space Science, Bucharest-Magurele, Romania
\item[$^{75}$] University Politehnica of Bucharest, Bucharest, Romania
\item[$^{76}$] Center for Astrophysics and Cosmology (CAC), University of Nova Gorica, Nova Gorica, Slovenia
\item[$^{77}$] Experimental Particle Physics Department, J.\ Stefan Institute, Ljubljana, Slovenia
\item[$^{78}$] Universidad de Granada and C.A.F.P.E., Granada, Spain
\item[$^{79}$] Instituto Galego de F\'\i{}sica de Altas Enerx\'\i{}as (IGFAE), Universidade de Santiago de Compostela, Santiago de Compostela, Spain
\item[$^{80}$] IMAPP, Radboud University Nijmegen, Nijmegen, The Netherlands
\item[$^{81}$] Nationaal Instituut voor Kernfysica en Hoge Energie Fysica (NIKHEF), Science Park, Amsterdam, The Netherlands
\item[$^{82}$] Stichting Astronomisch Onderzoek in Nederland (ASTRON), Dwingeloo, The Netherlands
\item[$^{83}$] Universiteit van Amsterdam, Faculty of Science, Amsterdam, The Netherlands
\item[$^{84}$] University of Groningen, Kapteyn Astronomical Institute, Groningen, The Netherlands
\item[$^{85}$] Case Western Reserve University, Cleveland, OH, USA
\item[$^{86}$] Colorado School of Mines, Golden, CO, USA
\item[$^{87}$] Department of Physics and Astronomy, Lehman College, City University of New York, Bronx, NY, USA
\item[$^{88}$] Louisiana State University, Baton Rouge, LA, USA
\item[$^{89}$] Michigan Technological University, Houghton, MI, USA
\item[$^{90}$] New York University, New York, NY, USA
\item[$^{91}$] Pennsylvania State University, University Park, PA, USA
\item[$^{92}$] University of Chicago, Enrico Fermi Institute, Chicago, IL, USA
\item[$^{93}$] University of Delaware, Department of Physics and Astronomy, Bartol Research Institute, Newark, DE, USA
\item[$^{94}$] University of Wisconsin-Madison, Department of Physics and WIPAC, Madison, WI, USA
\item[] -----
\item[$^{a}$] Fermi National Accelerator Laboratory, Fermilab, Batavia, IL, USA
\item[$^{b}$] Max-Planck-Institut f\"ur Radioastronomie, Bonn, Germany
\item[$^{c}$] School of Physics and Astronomy, University of Leeds, Leeds, United Kingdom
\item[$^{d}$] Colorado State University, Fort Collins, CO, USA
\item[$^{e}$] now at Hakubi Center for Advanced Research and Graduate School of Science, Kyoto University, Kyoto, Japan
\item[$^{f}$] also at University of Bucharest, Physics Department, Bucharest, Romania
\end{description}

\section*{The Telescope Array Collaboration}
\small

\begin{wrapfigure}[7]{l}{0.2\linewidth}
\vspace{-5mm}
\includegraphics[width=1.2\linewidth]{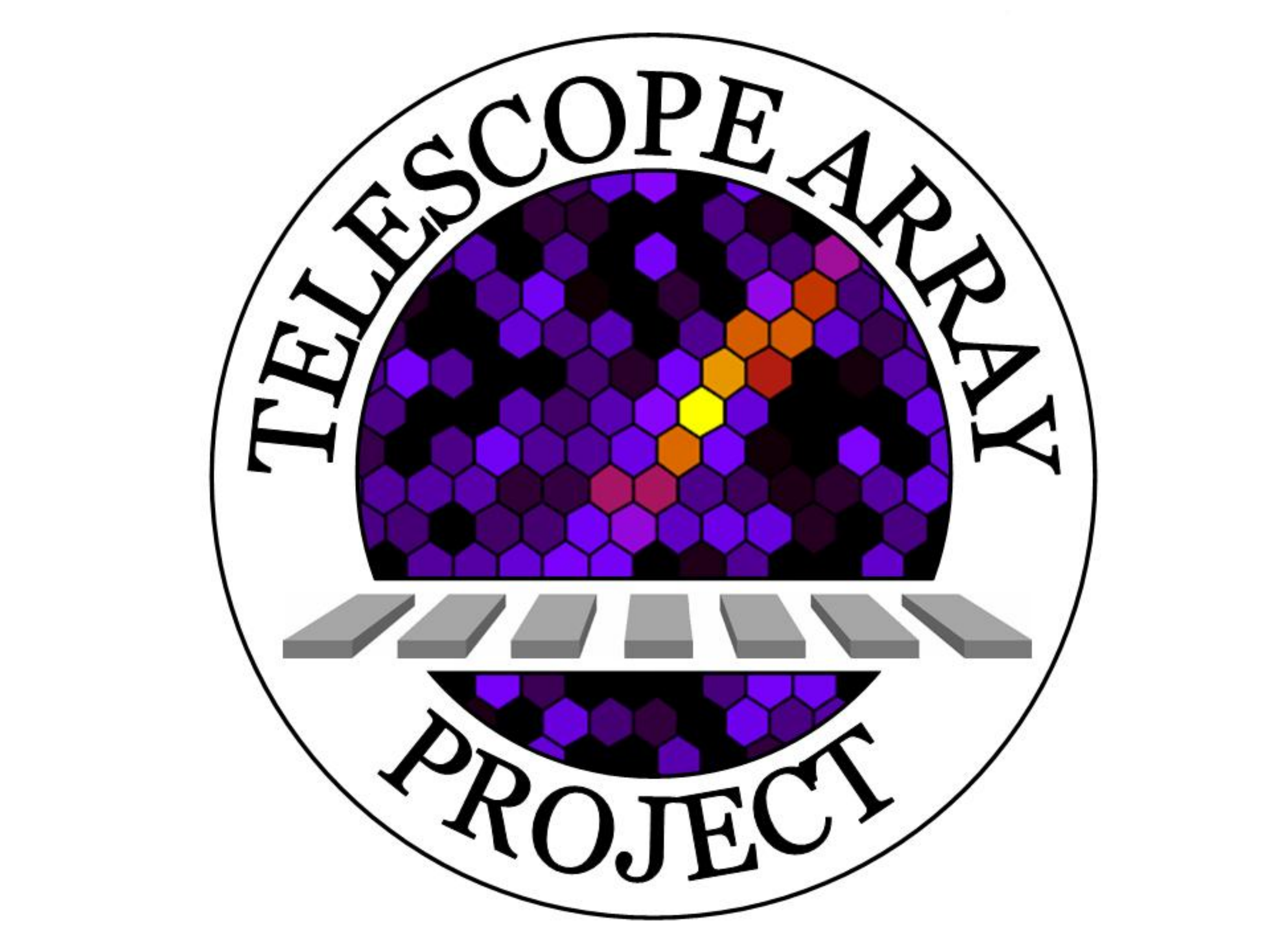}
\end{wrapfigure}
\begin{sloppypar}\noindent
R.U.~Abbasi$^{1,2}$,
T.~Abu-Zayyad$^{1,2}$,
M.~Allen$^{2}$,
Y.~Arai$^{3}$,
R.~Arimura$^{3}$,
E.~Barcikowski$^{2}$,
J.W.~Belz$^{2}$,
D.R.~Bergman$^{2}$,
S.A.~Blake$^{2}$,
I.~Buckland$^{2}$
% ...

\end{sloppypar}

\begin{center}
\rule{0.1\columnwidth}{0.5pt}
\raisebox{-0.4ex}{\scriptsize$\bullet$}
\rule{0.1\columnwidth}{0.5pt}
\end{center}

\vspace{-1ex}
\footnotesize

{\footnotesize\it
$^{1}$ Department of Physics, Loyola University Chicago, Chicago, Illinois, USA \\
$^{2}$ High Energy Astrophysics Institute and Department of Physics and Astronomy, University of Utah, Salt Lake City, Utah, USA \\
$^{3}$ Graduate School of Science, Osaka City University, Osaka, Osaka, Japan \\
$^{4}$ Department of Physics and The Research Institute of Natural Science, Hanyang University, Seongdong-gu, Seoul, Korea \\
% ...
}

\end{document}